\documentclass[journal]{IEEEtran}

\usepackage{amsmath,amssymb,amsfonts}
\usepackage{bm}
\usepackage{graphicx}
\usepackage[caption=false,font=footnotesize]{subfig} 
\usepackage{cite}
\usepackage{url}
\usepackage{algorithm}
\usepackage{algpseudocode}
\usepackage{thmtools}
\usepackage[hidelinks]{hyperref} 
\usepackage{amsmath}

\declaretheoremstyle[
  spaceabove=\topsep, spacebelow=\topsep,
  headfont=\normalfont\bfseries,
  notefont=\mdseries, notebraces={(}{)},
  bodyfont=\normalfont,
  postheadspace=1em,
  qed=\qedsymbol
]{mythmstyle}

\declaretheoremstyle[
  spaceabove=\topsep, spacebelow=\topsep,
  headfont=\normalfont\bfseries,
  notefont=\mdseries, notebraces={(}{)},
  bodyfont=\normalfont,
  postheadspace=1em,
  qed=$\diamond$
]{myremarkstyle}

\begin{document}

\allowdisplaybreaks
\pagenumbering{arabic}
\sloppy

\title{Communication in a Fractional World: MIMO MC-OTFS Precoder Prediction}

\author{ Evan Allen, Karim Said,  Robert Calderbank, and Lingjia Liu

\thanks{E. Allen, K. Said and L. Liu are with Wireless $@$ Virginia Tech, Bradley Department of Electrical Engineering, Virginia Tech. The corresponding author is L. Liu (ljliu@vt.edu).}}

\maketitle

\begin{abstract} 
As 6G technologies advance, international bodies and regulatory agencies are intensifying efforts to extend seamless connectivity especially for high-mobility scenarios such as Mobile Ad-Hoc Networks (\textit{MANETs}) types such as  Vehicular Ad-Hoc Networks (\textit{VANETs}) and Flying Ad-Hoc Networks (\textit{FANETs}). For these environments to be considered for long term adoption and use they must support Multiple-Input-Multiple- (MIMO) technology, rapidly fluctuating channel conditions in these environments place a heavy burden on traditional time-frequency CSI feedback schemes required for MIMO precoding. This motivates a shift toward delay-Doppler representations like those employed by Orthogonal Time-Frequency Space(OTFS) modulation, which offers greater stability under mobility. We derive an expression for the variation over time in the OTFS I/O relationship. We then use this to create a physics informed complex exponential basis expansion model prediction framework that maximizes the usefulness of outdated Channel State Information (CSI) in the presence of integer and fractional delay-Doppler channels and facilitates high mobility MIMO communication.

\end{abstract}


\section{Introduction}
Although Orthogonal Frequency-Division Multiplexing(OFDM) acts as the dominant waveform for modern digital communications thanks to its spectral efficiency and robust multi-path performance it encounters substantial difficulty in high velocity communications such as those seen in Mobile Ad-Hoc Networks (MANETs) and Non-Terrestrial Networks (NTN). MIMO-OFDM systems must contend with Doppler spread induced Inter-Carrier Interference (ICI) and balance signaling overhead against the accuracy and frequency of reporting channel state information. Under rapid channel fluctuations consistent with these scenarios, traditional OFDM struggles to enable effective communications \cite{Wang_2006}.

Stemming from these limitations, considerable research employs Orthogonal Time-Frequency Space (OTFS) modulation as a means of fortifying MIMO links in the presence of high mobility impairments. 
OTFS is known to exhibit outstanding performance in the presence of large delay-Doppler spreads while maintaining performance similar to that of OFDM in traditional communications environments \cite{Hadani_2017,Raviteja_2019}.   Much of the resilience of OTFS  is derived from its use of the delay-Doppler domain channel representation rather than the time-frequency channel for estimation and data transmission \cite{OTFS_BEM_EST}. The channel I/O relationship provided by OTFS is a reflection of the physical scattering environment and as such changes slower than the Time-Frequency representation \cite{Zak_Predictability}.  it is of course still subject to fundamental challenges and complexities that arise in high mobility environments as the physical environment develops.

An integral aspect to provide the coverage and throughput improvements set out by 5G and 6G systems is the use of MIMO beamforming by which signal energy can be shaped to allow for the effective separation of multiple data streams simultaneously.  These gains are dependent on accurate knowledge of the channel state information (CSI) at the transmitter, which is generally reported from user equipment (UE) periodically. It is natural then that as UEs experience greater velocity the reported feedback CSI is affected by inaccurate estimation and reduced coherence times. Mismatch between the true channel and that utilized for precoding can then lead to significant performance degradation, possibly leading to a bottleneck for the overall system performance \cite{Li_2006,Zhu_2012}.

We address the challenges implicit in OFDM CSI feedback by developing model based prediction of the interaction between the delay-Doppler domain channels and OTFS modulation. Thereby, we improve performance of MIMO precoding through the use of prediction without necessitating additional overhead required to improve performance such as increased reporting frequency of CSI. 

The literature offers extensive study and partial remedies of CSI feedback delay in OFDM systems  \cite{Li_2006}; nevertheless, most contributions concentrate on deficiencies unconsidered prior to modern developments in high mobility communication environments such as Flying Ad‐Hoc Networks (\textit{FANET}). Under high-velocity conditions, the detrimental effect of feedback latency becomes markedly more severe  \cite{NTNFeedback}. While the baseline behavior of OTFS has been studied for static and mobile channels, these investigations overlook the time-varying nature of OTFS links and the attendant impact of outdated feedback \cite{NTNFeedback,Hadani_2017,Raviteja_2019}. 

Other works have explored reinforcement learning for OTFS detection \cite{2D-RC} and machine learning-based precoder prediction for URLLC scenarios \cite{OTFS_ML_Pred}. In the case of OTFS detection, reinforcement learning benefits from pilot symbols as clear reference points, which simplifies the learning of accurate symbol decisions. By contrast, precoder prediction relies on synthetically generated, model-dependent channel realizations together with analytically computed SINR values to train the network. While this approach eliminates the need for labeled precoder data, it introduces greater initial training complexity and may exhibit limited generalization, necessitating retraining as actual channel conditions change.
Prior work in utilization of basis expansion modeling (BEM) for OTFS focuses around its use in channel estimation and its application to the time domain rather than its use in the delay-Doppler domain. \cite{OTFS_BEM_EST}. Closest to our applications \cite{OTFS_BEM_EST} considers an alternative form of Discrete-Prolate-Spheroidal (DPS) - BEM for prediction of DL channels from a delayed UL estimation. This work is different than ours through its consideration of an alternative BEM modeling that does not leverage the intuitive frequency estimation present in OTFS, further they do not demonstrate MIMO precoding performance rather they provide a calculated sum rate. Further they do not provide comparison of performance under OTFS and OFDM.  

This work is an extended version of our prior conference paper presented at VTC2024-Fall \cite{VTC_Prior}. In this journal version,  we present an analysis of  fractional delay and Doppler where paths and their resultant channel behavior.  To address these issue, we propose an adjusted method for prediction in fractional channels. We further provide results for estimated integer channels to compare the resiliency of channel estimation for each modulator in high Doppler environments.  We then evaluate performance across a range of mobility centric scenarios relevant to the broader research community. 
The contributions in this work include:
 \begin{itemize}
    \item Developed a delay-Doppler domain BEM predictive framework for prediction in the presence of heavy fractional paths, including performance under idealized and impaired channel sensing. 
    \item Considerations for channel estimation on integer OTFS prediction. 
    \item Monte Carlo simulation showcasing the comparable OFDM and OTFS performance in integer and fractional channels across mobility and estimation errors. 
\end{itemize}

The paper is then organized as follows:
Section \ref{se:System} covers the fundamental channel model and system configurations considered as well as any assumptions implicit therein. Section \ref{se:Prediction} then delves into the expression and concepts related to our OTFS based prediction as well as the OFDM based prediction to which we compare results. Section \ref{se:Results} \& \ref{se:conclusion} then describes the simulation environments and considerations paired followed by a summary of our key contributions, insights and future research direction. 


\section{System Model}
\label{se:System}
\subsection{OTFS Modulation}
Orthogonal Time-Frequency Space (OTFS) modulation consists of the mapping of modulated symbols onto a two-dimensional lattice (hereafter referred to as a grid), analogous to assigning symbols to the
\( N_{\text{subcarriers}} \!\times\! N_{\text{symbols}}\) resource grid in OFDM. Each grid symbol is subsequently processed by a two-dimensional OTFS modulation kernel that spreads the symbol energy across a delay-Doppler grid characterized by N delay bins and M Doppler bins.

In this work we adopt the multi-carrier variant, MC-OTFS, whose transmit signal is generated via the Heisenberg transform described in~\cite{Hadani_2017}.

\begin{equation}
  s(t) \;=\; 
  \sum_{m=0}^{M-1} \sum_{n=0}^{N-1}
  X[n,m] \, e^{j 2\pi m \,\Delta f \,(t - nT)} \,
  g_{\mathrm{tx}}\!\bigl(t - nT\bigr)
\end{equation}

When considering ideal pulses, 
\[
g_{\mathrm{tx}}(t)
=
\frac{1}{\sqrt{T}}\,
\operatorname{rect}\!\Bigl(\tfrac{t}{T}\Bigr)
=
\begin{cases}
1, & 0 \le t < T,\\[4pt]
0, & \text{otherwise},
\end{cases}
\]
The signal \textit{s(t)} is then acted upon by a channel matrix \textbf{H} of size \(MN\!\times\!MN\) where $M,N$ represent the number of delay and Doppler bins respectively. Choosing delay spacing  $\frac{1}{M \Delta f}$ and Doppler spacing $\frac{\Delta f}{N}$ yields an MN-dimensional orthonormal DD basis \cite{OTFS_Principals}.

\subsection{The Wireless Channel}
We begin with a general continuous-time convolution model using the time-varying impulse response \( h(t,\tau) \), parameterized by complex path gains \( h_p \), delays \( \tau_p \), and Doppler shifts \( \nu_p \):

\begin{equation}
  y(t) = \int_{-\infty}^{\infty} 
  h(t,\tau)\,x\!\bigl(t-\tau\bigr)\,d\tau + n(t)
\end{equation}

\begin{equation}
  h(t,\tau) = \sum_{p=0}^{P-1} h_{p} \, \delta\!\bigl(\tau-\tau_{p}\bigr)\, e^{j 2\pi \nu_{p} t}
\end{equation}

Given the finite bandwidth of the signal $\frac{1}{T_s}$, we discretize the model using delay \( k \) and Doppler  \( m \), where \( t = n_s T_s \) and \( T_s = \frac{1}{\Delta f} \). The discrete-time received signal  for the jth sample becomes:

\[
y[j]  =\sum_{k=0}^{N-1}  h[j,k] \, x[ k]  + n[j]
\]

where

\[
h[j,k] = \sum_{p=0}^{P-1} h_p \, \text{sinc}(m-k-\tau_p/T_s) \, e^{j 2\pi \nu_p j T_s}
\]

We further develop this model acting on a transmission block \( \tilde{\mathbf{x}} = [\mathbf{c}^T, \mathbf{x}^T]^T \), where \( \mathbf{x} = [x_1, \dots, x_L]^T \) is a block of \( L \) data symbols and \( \mathbf{c} = [x_{L-c+1}, \dots, x_L]^T \) is a cyclic prefix of length \( c \). The term \textit{block} refers to either an OFDM data frame or a reduced-CP OTFS frame \cite{hong2022delay}.

The corresponding effective time-variant impulse response matrix that acts on a block of length \( L \), incorporating CP effects, is given by \cite{8516353}:

\begin{equation} \label{discrete_time_scattering_model}
\mathbf{H} = \sum_{p=0}^{P-1} h_p \, \Delta_{L}(\nu_p) \, \Pi_{L}(\tau_p)
\end{equation}

where each of $h_p, \tau_p, \nu_p$ represents the gain, normalized delay (in units of samples) and normalized Doppler shift (in units of cycles/block) of the $p$-th path respectively, $\Pi_{L}(\tau_p) \ \in \mathbb{C}^{L\times L}$ models the delay effect, and $\Delta_{L} \ \in \mathbb{C}^{L\times L}(\nu_p)$ models the Doppler shift effect respectively as defined by \eqref{delay_matrix}
\begin{equation}\label{delay_matrix}
\Pi_{L}(\tau_p)\triangleq \mathbf{F}_{L}\mathbf{D}_{L}(\tau_p)\mathbf{F}_{L}^H, \quad\Delta_{L}(\nu_p)\triangleq \mathbf{D}_{L}(\nu_p)
\end{equation}

where $\mathbf{D}_{L}(x) \in \mathbb{C}^{L\times L}$ is a diagonal matrix with elements $\lbrace\mathbf{D}_{L}(x)\rbrace_{r,c}=z_L^{xr}\delta(r-c) \quad r,c=0,..,L-1$, $\lbrace\mathbf{F}_{L}\rbrace_{r,c}=\frac{1}{\sqrt{L}}z_L^{rc} \quad r,c=0,..,L-1$ and $z_L=e^{j\frac{2\pi}{L}}$.

For a choice of CP length $c\geq \tau_{max}$, the  continuous channel is segmented into channel effects on a  sequence of consecutive blocks $\tilde{\mathbf{x}}_i,\tilde{\mathbf{x}}_{i+1},..$  where the $i$-th block channel is given by:

\begin{equation}\label{per_block_channel}
\mathbf{H}[i]=\sum_{p=0}^{P-1}h_p z_L^{i(c+L)\nu_p}\Delta_{L}(\nu_p)\Pi_{L}(\tau_p)
\end{equation}
the term $z_{L}^{(c+L)\nu_p}$ accounts for time by adding a time dependent  phase progression term correspondent to the Doppler (normalized with respect to block size $L$) of the $p$-th path.

To effectively compare OFDM and OTFS, we unify the parameter values such that a block carries $MN$ data symbols for both systems. We define an an OFDM frame as a sequence of $M$ blocks each having effective length $N$ (number of sub-carriers), making the total frame length $N(M+c)$ samples. An OTFS frame is defined as a single block of effective length $dMN$, i.e., a total frame length $c+MN$. Similarly, we define. For the purposes of the per-block channel given in \eqref{per_block_channel}, an OFDM block channel acts on one OFDM symbol and an OTFS block channel acts on a full OTFS frame.
These selections ensure equal data transfer while allowing OTFS to utilize a satisfactory delay-Doppler grid resolution.

For a sequence of OFDM frames, the channel evolution on an OFDM symbol level is given by:
\begin{equation}\label{per_block_channel_OFDM}
\mathbf{H}^F[s]=\sum_{p=0}^{P-1}h_p z_{MN}^{s(c+N)\nu_p}\Delta_{N}\left(\frac{\nu_p}{M}\right)\Pi_{N}(\tau_p)
\end{equation}
where $s=0,1,...$ is the symbol index, and the normalized Doppler utilized by the phase progression term
\begin{align*}
    z_{MN}^{s(c+N)\nu_p}
\end{align*}
is normalized with respect to the effective frame size $MN$, i.e., $\nu_p \in \mathbb{R}$  is in units of number of cycles completed over delay-Doppler samples.

For a sequence of OTFS frames, the channel evolution across frames is denoted by $\mathbf{H}^O[r]$
\begin{equation}\label{per_block_channel_OTFS}
\mathbf{H}^O[r]=\sum_{p=0}^{P-1}h_p z_{MN}^{r(MN+c)\nu_p}\Delta_{MN}(\nu_p)\Pi_{MN}(\tau_p)
\end{equation}
where $r$ is the OTFS frame index.

\subsection{OFDM equivalent channel matrix}
Under the integer delay and integer Doppler assumption we rewrite \eqref{per_block_channel_OFDM} in the form given by \eqref{per_block_channel_OFDM_int_delay_Dop}, where the sum is over all possible integer delay and Doppler values, represented by  $n,m$, respectively. The channel gains denoted by $h_{n,m}$, are non-zero only for the $(n,m)$ pairs contained in the set $\lbrace \lbrace0,..,\tau_{max}\rbrace \times \lbrace M-\nu_{max},...,M-1 \rbrace \cup \lbrace 0,..,\nu_{max}\rbrace$
\begin{equation}\label{per_block_channel_OFDM_int_delay_Dop}
\begin{split}
\mathbf{H}^F
&=\sum_{n=0}^{N-1}\sum_{m=0}^{M-1}h_{n,m} z_{N}^{sm(c+N)}\Delta_{N}\left(\frac{m}{N}\right)\Pi_{N}^n\\
\end{split}
\end{equation}  
where $\Pi_{N}\triangleq \Pi_{N}(1)$ is a cyclic shift matrix , $\tau_{max} $ , $\nu_{max} $ are the maximum delay and Doppler respectively, and we get the simplification $\lbrace\Pi_{N}(n)\rbrace_{p,q}=\delta(\langle p-q-n\rangle_N)$. We note that $\Pi_{N}^n$
is $\Pi_{N}$ raised to power $n$.

Now we find the per-symbol equivalent OFDM channel matrix, denoted by $\mathbf{H}_{eq}^F$, which is the result of applying receiver and transmit side discrete Fourier transform(DFT) operations on a per symbol basis :

\begin{equation}\label{per_block_channel_OFDM_int_delay_Dop_eq_ch}
\begin{split}
\mathbf{H}_{eq}^F[s]&=\mathbf{F}_N\mathbf{H}^F[s]\mathbf{F}_N^H\\
&=\sum_{n=0}^{N-1}\sum_{m=0}^{M-1}h_{n,m} z_{N}^{sm(c+N)}\Pi_{N}\left(\frac{m}{M}\right)\mathbf{D}_N^n\\
\end{split}
\end{equation}  
where $\mathbf{D}_N\triangleq \mathbf{D}_N(1)$.
We assume symbol-level channel quasi-staticity. This facilitates a time-invariant (TI) channel on a per-symbol basis which is typical for OFDM \cite{OFDM1} ,\cite{OFDM2} ,\cite{OFDM3}, \cite{OFDM4}
\begin{equation}\label{per_block_channel_OFDM_TI_channel}
\begin{split}
\mathbf{H}_{eq}^F[s]&\approx\sum_{n=0}^{N-1}\tilde{h}_n(s)\mathbf{D}_N^n\triangleq \mathbf{H}_{eq}^{F_{TI}}[s]
\end{split}
\end{equation}  
where $\tilde{h}_n(s)=\sum_{m=0}^{M-1}h_{n,m} z_{MN}^{sm(c+N)}\Pi_{N}\left(\frac{m}{M}\right)$, which is the $n$-th delay tap for the $s$-th OFDM symbol, $\mathbf{H}_{eq}^{F_{TI}}$ is a diagonal matrix representing the per-subcarrier gain channel due to the TI channel approximation 

The Doppler effect introduces a frequency offset between the transmitted and received signals due to relative motion. defining the transmit signal s(t) by subcarrier \( k \), this offset can be correspondingly expressed as:
\begin{equation}
s(t)= \frac{1}{\sqrt{T}} \sum_{k=0}^{K-1} a_k \, e^{j 2 \pi k t} \, w(t) \quad k \in \{0, \ldots, K - 1\}
\end{equation}

\[
f_\delta = f_{\text{Tx}} - f_{\text{Rx}},
\]
where \( f_{\text{Tx}} \) is the carrier frequency of the transmitter and \( f_{\text{Rx}} \) is the frequency perceived at the receiver. This offset manifests as a phase rotation in the received signal, which can be modeled as:

\[
y(t) = s(t) e^{j 2 \pi f_\delta t}
\]
where \( s(t) \) is the transmitted baseband signal and \( y(t) \) is the received signal after experiencing the frequency shift \( f_\delta \).

We normalize the frequency offset by the subcarrier spacing \( \Delta f \), yielding:
\begin{equation}
y(t) = s(t) e^{j 2 \pi \frac{f_\delta}{\Delta f} t}
\end{equation}
where \( \delta \) is the normalized frequency offset (in subcarrier units).

The demodulated signal on the \( k \)-th subcarrier is obtained by correlating the received signal \( y(t) \) with the basis function of subcarrier \( \ell \). This is given by:
\begin{equation}
r(k) = \frac{1}{\sqrt{T}} \int_0^T y(t) \cdot e^{-j 2 \pi \frac{\ell t}{\Delta f}} \, dt,
\end{equation}
which simplifies, using the expression for \( y(t) \), to:
\begin{equation}
r(k) = \frac{1}{\sqrt{T}} \int_0^T s(t) \cdot e^{j 2 \pi \frac{(k + \delta - \ell)t}{T}} \, dt, \quad \ell \in \{0, \ldots, K - 1\}.
\end{equation}

Assuming that the transmitted signal \( s(t) \) remains approximately constant over the duration of an OFDM symbol (a reasonable assumption in flat-fading channels), the presence of fractional Doppler shifts causes energy to leak between adjacent subcarriers. This leakage follows a sinc-like pattern, and in the discrete domain, it is approximated by the Dirichlet kernel. The resulting expression for the received signal on subcarrier \( n \), accounting for Doppler-induced leakage from a neighboring subcarrier \( \ell \), is given by:
\begin{equation}
r(n) \approx s(n) \cdot  \frac{\sin\left(\pi N (n - \ell - \delta_m)\right)}{N \sin\left(\pi (n - \ell - \delta_m)\right)} , \quad \ell \in \{0, \ldots, K - 1\}.
\end{equation} 

\begin{equation}
[\mathbf{C}_{\text{ICI}}]_{k,\ell} = \frac{\sin\left(\pi N (k - \ell - \delta_m)\right)}{N \sin\left(\pi (k - \ell - \delta_m)\right)}
\end{equation}

This matrix characterizes how energy spreads from subcarrier \( \ell \) to subcarrier \( n \) due to the loss of orthogonality induced by Doppler shift. This is then representative of inter carrier interference known to affect OFDM systems in the presence of Doppler shifts. Considering this our effective OFDM I/O relationship becomes. 
\begin{equation}\label{per_block_channel_OFDM_frac_Doppler}
\begin{split}
\mathbf{H}_{eq}^F[s] &= \mathbf{F}_N \mathbf{H}^F[s] \mathbf{F}_N^H \\
&= \sum_{n=0}^{N-1} \sum_{m=0}^{M-1} h_{n,m} \, z_{MN}^{sm(c+N)} \, \mathbf{C}_{\text{ICI}}\left(\tfrac{m}{M}\right) \, \mathbf{D}_N^n
\end{split}
\end{equation}

\subsection{OTFS equivalent channel matrix}
We start from \eqref{per_block_channel_OTFS} adapted to the integer delay and Doppler assumption:
\begin{equation}\label{per_frame_channel_OTFS_int_delay_Dop}
\begin{split}
\mathbf{H}^O[r]&=\sum_{n=0}^{N-1}\sum_{m=0}^{M-1}h_{m,n} z_{MN}^{r(MN+c)m}\Delta_{MN}^m\Pi_{MN}^n\\
&=\sum_{n=0}^{N-1}\sum_{m=0}^{M-1}h_{m,n} z_{MN}^{rcm}\Delta_{MN}^m\Pi_{MN}^n
\end{split}
\end{equation} 
To obtain the equivalent channel matrix form, we left and right multiply \eqref{per_frame_channel_OTFS_int_delay_Dop}
by unitary OTFS modulation matrices $\mathbf{O}$ and $\mathbf{O}^H$ respectively:
\begin{equation}\label{per_frame_eq_channel}
\begin{split}
\mathbf{H}_{eq}^O[r]&=\mathbf{O}\mathbf{H}^O[r]\mathbf{O}^H\\
&=\sum_{n=0}^{N-1}\sum_{m=0}^{M-1}h_{m,n} z_{MN}^{rcm}\mathbf{O}\Delta_{MN}^m\Pi_{MN}^n\mathbf{O}^H
\end{split}
\end{equation} 
where $\mathbf{O}=\mathbf{F}_M \otimes \mathbf{I}_N$ is the OTFS modulation matrix, the Kronecker product of $M$ the length of the Doppler dimension and $N$ is the length of the delay dimension.\cite{8516353}.

Eq. \eqref{per_frame_eq_channel} can be succinctly transformed into a more compact expression as follows. It's important to note that the derivation of this reduced form is detailed in prior works and has been omitted from the present document due to space constraints. 
\begin{equation}\label{per_frame_eq_channel_reduced_form}
\begin{split}
\mathbf{H}_{eq}^O[r]&=\mathbf{O}\mathbf{H}^O[r]\mathbf{O}^H\\
&=\sum_{n=0}^{N-1}\sum_{m=0}^{M-1}h_{m,n} z_{MN}^{(r+1)cm}\mathbf{P}_m\left(\Pi_{M}^{-m} \otimes\Pi_{N}^n\right)\mathbf{Q}_n^H
\end{split}
\end{equation} 
where $\mathbf{P}_m$ and $\mathbf{Q}_n$ are diagonal matrices by \eqref{P_m} and \eqref{Q_n} respectively.
\begin{equation}\label{P_m}
\mathbf{P}_m = \mathbf{I}_M \otimes \textbf{diag}(\mathbf{x}_m), \lbrace\mathbf{x}_m\rbrace_{c} = e^{j2\pi \frac{mc}{MN}} 
\end{equation}
\begin{equation}\label{Q_n}
\lbrace\mathbf{Q}_n\rbrace_{r,c} = \delta(r-c) 
\\
\exp\left(j2\pi\chi_{\left(\langle c\rangle_N > N-n\right)}\frac{\lfloor\frac{c}{N}\rfloor }{M}  \right) \\
\end{equation}

In scenarios where the resolution of the delay-Doppler grid is insufficient to fully resolve fractional delay or Doppler shifts, the energy of a transmitted path is not confined to a single delay-Doppler bin. Instead, it spreads across neighboring bins. This spectral leakage effect can be mathematically characterized by a sinc-shaped function. In discrete systems, this is approximated by the Dirichlet kernel, which captures how the signal energy is distributed over adjacent delay or Doppler indices due to fractional bin misalignment. \cite{OTFS_Principals}
\[
\begin{array}{ll}
\tau_n = \frac{2\pi}{N}(i_n - p_n), & 
\tau_{\text{dirch}} = \frac{\sin\left(\frac{N \tau_n}{2}\right)}{N \sin\left(\frac{\tau_n}{2}\right)} \\[10pt]
\nu_m = \frac{2\pi}{M}(i_m - p_m), & 
\nu_{\text{dirch}} = \frac{\sin\left(\frac{M \nu_m}{2}\right)}{M \sin\left(\frac{\nu_m}{2}\right)}
\end{array}
\]
\begin{equation}\label{per_frame_eq_channel_dirch_form}
\begin{split}
\mathbf{H}_{eqFrac}^O[r] &= \mathbf{O} \mathbf{H}^O[r] \mathbf{O}^H \\
&= \sum_{n=0}^{N-1} \sum_{m=0}^{M-1} h_{m,n} \, z_{MN}^{(r+1)cm} \, \mathbf{P}_m \left( \mathbf{\nu}_{\text{dirch}}(m) \otimes \mathbf{\tau}_{\text{dirch}}(n) \right) \mathbf{Q}_n^H
\end{split}
\end{equation}
  
Eq.~\eqref{per_frame_eq_channel_reduced_form} illustrates the structure of the equivalent OTFS channel matrix under the assumption of sufficiently high delay-Doppler resolution such that all propagation paths exhibit integer delay and Doppler shifts. In this setting, the channel matrix is expressed as a weighted sum of per-path contributions. The matrices \( \mathbf{P}_m \) and \( \mathbf{Q}_n \) capture the effects of employing a rectangular pulse shape. If these matrices were removed-effectively replacing the rectangular window with an ideal Dirac pulse-the resulting structure would correspond to a perfect two-dimensional (2D) circular convolution.

In this analysis we consider ideal pulse shapes. Extending to rectangular, raised-cosine, and other pulses is an active area of study

\begin{figure}[h!]
\centering
    \includegraphics[width=1\columnwidth]{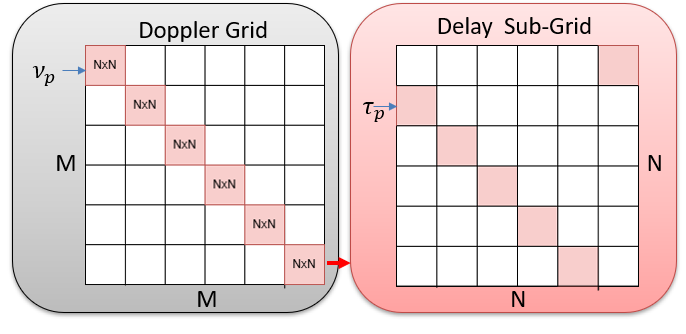}
    \caption{OTFS Effective Channel Block-wise structure: $MN \times MN$}  
  \textbf{Left:} Doppler Grid: $M \times M$ structure for an integer Doppler path 0.  
  \textbf{Right:} Delay Subgrid: $N \times N$ for an integer delay path 1.
  \label{Fig.IntegerDDChannel}
\end{figure}

\begin{figure}[h!]
\centering
    \includegraphics[width=1\columnwidth]{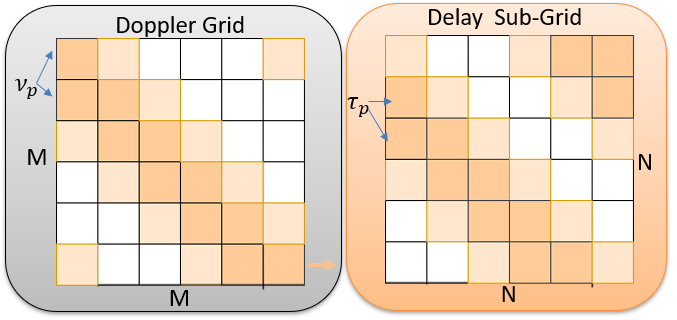}
    \caption{Effective OTFS Channel I/O representation with Fractional Spreading}  
  \textbf{Left:}Doppler Bin : $M \times M$ structure for an Fractional Doppler path .5.  
  \textbf{Right:} Delay subgrid: $N \times N$ for an integer delay path 1.5.
  \label{Fig.IntegerFractionalChannel}
\end{figure}
 Fig \ref{Fig.IntegerDDChannel} shows a graphic representation of the 2D circular convolution structure of the vectorized OTFS I/O representation. This is conceptually a matrix composed of submatrices. Within each submatrix(right) the delay path determines the occupied sections. This is repeated in the overall matrix(left) wherein the Doppler paths determine which submatrices are occupied.  

 Fig. \ref{Fig.IntegerFractionalChannel} shows the spreading of a fractional delay-Doppler path across multiple integer bins correspondent to the Dirichlet kernel spreading.  

In contrast, the equivalent channel I/O matrix for OFDM, as shown in Eq.~\eqref{per_block_channel_OFDM_TI_channel}, is comprised by weighted sum of cosines per-path. These contributions \textit{superimpose} in the received signal vector, leading to interference across basis functions. The greater capability of MC-OTFS overlays to minimize interference through greater DD path separability even under rich multipath conditions is fundamental to the performance advantages analyzed and leveraged in the subsequent sections of this work.

\subsection{Cyclic prefix condition for periodicity of OTFS channel matrix}

Following our OTFS effective channel \eqref{per_frame_eq_channel_reduced_form} the evolution across data frames is determined by the Doppler phase term $z_{MN}^{(r+1)cm}$which has the property: for
$c=kN$,  $k\in \mathbb{Z}$, $z_{delay-Doppler}^{(r+1)kNm}=z_{M}^{(r+1)km}$. As a result the Doppler phase term becomes periodic in $m$ with period $M$. 

Substituting in \eqref{per_frame_eq_channel_reduced_form}
\begin{equation}\label{per_frame_eq_channel_predictable_form}
\begin{split}
\mathbf{H}_{eq}^O[r]
&=\sum_{n=0}^{N-1}\sum_{m=0}^{M-1}h_{m,n} z_{M}^{(r+1)km}\mathbf{P}_m\left(\Pi_{M}^{-m} \otimes\Pi_{N}^n\right)\mathbf{Q}_n^H\\
&=\left(\mathbf{L}[r]\otimes \mathbf{I}_N\right)\mathbf{H}_{eq}^O[0]\left(\mathbf{L}[r]\otimes \mathbf{I}_N\right)^H\
\end{split}
\end{equation} 
where $\lbrace \mathbf{L}(r)\rbrace_{p,q} = e^{j2 \pi \frac{(r+1)pk}{M} } \delta(p-q), \quad p,q=0,..,M-1$.

Eq. \eqref{per_frame_eq_channel_predictable_form}  reveals the predictable nature of the channel across frames where the change across frames is due to a deterministic factor $\mathbf{L}(r)$. This finding serves as the basis for the prediction of the interaction between OTFS modulation and the integer delay-Doppler channel.

In the following sections we extend this analysis to the MIMO case and provide a procedure for direct precoder prediction. 

\subsection{Impulse‐Based Pilot Design in Delay-Doppler Domain}
OTFS naturally admits a very low‐overhead, high‐resolution channel‐estimation strategy by placing a single “impulse” pilot symbol in the delay-Doppler grid and allocating a guard region around \((k_p,\ell_p)\) defined by the maximum delay and Doppler spread to avoid pilot-data interference.  Concretely, let
\[
X_p[k,\ell] =
\begin{cases}
1, & (k,\ell) = (k_p,\ell_p),\\
0, & \text{otherwise},
\end{cases}
\]
  After transmission through the doubly‐dispersive channel, the received delay-Doppler grid is
\[
\begin{aligned}
Y_p[k,\ell]
&= \bigl(\mathbf{H}_{\rm DD}\,\star\,X_p\bigr)[k,\ell] + W[k,\ell],\\
&= \mathbf{H}_{\rm DD}\bigl[k-k_p,\;\ell-\ell_p\bigr]     + W[k,\ell].
\end{aligned}
\]
where \(\star\) denotes 2-D circular convolution, \(H_{\rm DD}[\,\cdot,\cdot\,]\) is the delay-Doppler channel impulse response, and \(W[k,\ell]\) is noise.  Hence, by simply reading off \(Y_p\) over the known support, one recovers all nonzero taps of \(H_{\rm DD}\) in one shot.

When multiple transmit antennas are sounded simultaneously, orthogonality can be preserved by assigning distinct pilot positions \((k_p,\ell_p)\) per antenna in the delay-Doppler grid~\cite{OTFS_Estimation}.  A simple least‐squares estimator then yields the DD-domain channel estimate with just \(O(P)\) complexity-where \(P\) is the number of significant paths-instead of a full \(O((MN)^2)\) solve.  This impulse‐pilot approach exploits both the inherent sparsity of wireless channels in delay-Doppler and the block‐circulant structure of OTFS, delivering high‐resolution estimates with minimal overhead.


\section{MIMO Precoder Design \& Prediction}
\label{se:Prediction}

\subsection{MIMO Considerations}
We consider a MIMO setup consisting of $N_t$ transmit antennas and $N_r$ receive antennas where $N_r\geq N_t$. 
We define a MIMO input symbols vector for OFDM as,
\begin{equation}\label{DataSymbols}
\bar{\mathbf{x}}^F=[\tilde{\mathbf{x}}_{0}^T,..,\tilde{\mathbf{x}}_{N-1}^T]^T
\end{equation}
where $\tilde{\mathbf{x}}_{n}\in \mathbb{C}^{N_t\times 1}$ consists of data in the $n$-th subcarrier, and $\bar{\mathbf{x}}^F \in \mathbb{C}^{NN_t\times 1}$ represents data encoded in $N_t$ spatial streams within one OFDM symbol.

For OTFS, MIMO transmission is to be done on a frame level, where the input symbol vector is
\begin{equation}\label{MIMO_OFDM}
\bar{\mathbf{x}}^O=[\mathbf{x}_{0}^T,..,\mathbf{x}_{N_t-1}^T]^T
\end{equation}
where $\mathbf{x}_t \in \mathbb{C}^{MN\times 1}$.

The MIMO channel matrix for the $s$-th OFDM symbol is denoted by a block diagonal matrix $\bar{\mathbf{H}}^F[s]\in \mathbb{C}^{NN_r\times NN_t }$ 
\begin{equation}\label{MIMO_OFDM}
  \bar{\mathbf{H}}^F[s] = \textbf{blkdiag}\left(  \tilde{\mathbf{H}}_0[s],\hdots,\tilde{\mathbf{H}}_{N-1}[s] \right)    
\end{equation}
where $\tilde{\mathbf{H}}_l[s] \in \mathbb{C}^{N_r\times N_t}$, and $\lbrace\tilde{\mathbf{H}}_l[s]\rbrace_{p,q}=\sum_{n=0}^{N-1}\tilde{h}_n^{p,q}(s)\lbrace\mathbf{D}_N^n\rbrace_{l}$ is the $l$-th subcarrier gain for the $p$-th receiver, $q$-th transmitter channel where $\tilde{h}_n$ is defined in \eqref{per_block_channel_OFDM_TI_channel}. 

For OTFS, the MIMO channel matrix for the $r$-th frame is denoted by  $\bar{\mathbf{H}}^O[r]\in \mathbb{C}^{MNN_r\times MNN_t }$ 

\begin{equation}\label{MIMO_OTFS}
  \bar{\mathbf{H}}^O[r] =
  \begin{bmatrix}
    \mathbf{H}^O_{0,0}[r] & ... & \mathbf{H}^O_{0,N_t-1}[r] \\
    & \ddots & \\
    \mathbf{H}^O_{N_r-1,0}[r]&... & \mathbf{H}^O_{N_r-1,N_t-1}[r] 
  \end{bmatrix}
\end{equation}
where 

$\mathbf{H}^O_{p,q}[r]=\sum_{n=0,m=0}^{N-1,M-1}h^{p,q}_{m,n} z_{MN}^{(r+1)cm}\mathbf{P}_m\left(\Pi_{M}^{-m} \otimes\Pi_{N}^n\right)\mathbf{Q}_n^H$.

\subsection{Precoder Generation}
We adopt Zero-Forcing precoding as our baseline because of its straightforward implementation and proven reliability. 

Due to the block diagonal structure of the OTFS MIMO effective channel, transmit precoders are generated on a per delay bin basis where the precoder for the l-th delay bin is $\mathbf{W}_{l} =\mathbf{H}^H \bigl(\mathbf{H} \mathbf{H}^H\bigr)^{-1}$

For OTFS, the precoder is obtained for the entire frame by applying ZF to the matrix $\mathbf{H}^O_{0,0}[r]$. Although this incurs significant computational complexity compared to the OFDM case, the payoff will be a reduced feedback overhead due to the precoder predictability for OTFS which we derive in the following section.   

\subsection{Integer OTFS Precoder Prediction }

Substituting in \eqref{MIMO_OTFS} using \eqref{per_frame_eq_channel_reduced_form}

\begin{equation}
\bar{\mathbf{H}}^O_{0,0}[r]=\mathbf{{\bar{L}}}_{N_r}[\mathbf{r}]\bar{\mathbf{H}}^O_{0,0}[0]\mathbf{{\bar{L}}}_{N_t}[\mathbf{r}]
  \end{equation}
where
\begin{equation}
\mathbf{{\bar{L}}}_{Nr}[r]  = \mathbf{I}_{N_r} \otimes (\mathbf{L}[r] \otimes \mathbf{I}_{N} ),\quad  \mathbf{{\bar{L}}}_{Nr}[t] = \mathbf{I}_{N_t} \otimes (\mathbf{L}^{H}[r] \otimes \mathbf{I}_{N} )
\end{equation}


Therefore, we can see that $r$-th frame MIMO channel can be derived given knowledge of the $0$-th frame and the deterministic factors $\mathbf{{\bar{L}}}_{Nr}[r]$ and $\mathbf{{\bar{L}}}_{Nt}[r]$.

Finally, applying this to ZF operations we can find prediction equations for the linear precoder as follows given that $\mathbf{{\bar{L}}}_{Nr}[r], \mathbf{{\bar{L}}}_{Nt}[r]$ are unitary matrices:
\begin{equation}
\mathbf{W}[r] = \mathbf{{\bar{L}}}_{Nr}^H[r] W[0]\mathbf{{\bar{L}}}_{Nt}^H[r] 
\end{equation}

Therein we can utilize factors $\mathbf{{\bar{L}}}_{Nr}[r]$ and $\mathbf{{\bar{L}}}_{Nt}[r]$ in order to directly update the OTFS MIMO precoders. 

\subsection{Complex Exponential Basis Expansion}
Recall that we model the channel as:
\[
h[j,k] = \sum_{p=0}^{P-1} h_p \, \text{sinc}(m-k-\tau_p/T_s) \, e^{j 2\pi \nu_p j T_s}
\]

$h[j,k]$ is a sum of complex exponential spread by a sinc pulse with respect to delay $k$ and Doppler $m$. As this sum is nonlinear with respect to the time over j samples. We then select a estimator reflective of this structure.  We then select Complex Exponential Basis‐Expansion Model (CE‐BEM)~\cite{CE_BEM}. CE-BEM was selected for its fundamental structure for estimating a sum of complex sinusoids. In CE‐BEM the time‐varying channel impulse response is approximated by a finite sum of $Q=2N+1$ complex exponential, where $N$ is the number of estimated paths within the channel:
\[
  h(t)\;\approx\;\sum_{q=0}^{Q-1}c_qe^{j2\pi\,f_q T_{s}}
\]

The complex weights $\{c_q\}$ are obtained via a least‐squares estimate from prior time samples, and the Doppler shifts $\{f_q\}$ are drawn uniformly from the interval:
\[
  f_q\;\sim\;\mathcal{U}\!\Bigl[-\tfrac{D_{\max}}{2},\,\tfrac{D_{\max}}{2}\Bigr]\,,
\]
which yields a Doppler resolution between basis functions of:
\[
  \Delta f \;=\; \frac{2F_D}{Q}\,.
\]
We then repeat this estimation on a per subcarrier basis where a sliding window of previous channel estimates across symbols is utilized to determine $c_q$ and predict the future subframe.

This estimation is then reliant upon Q's span encompassing the maximum Doppler shift and the ability to effectively estimate the channel in a high mobility environment for the calculation of $c_q$.

\subsubsection{Fractional OTFS Prediction }
To consider fractional OTFS precoder prediction the diagonal structure of the 2D circulation matrix experiences spreading proportional to the fractional elements of delay and Doppler. This means that at a given delay-Doppler bin there now exists a summation of complex exponential rather than a unique diagonal path \ref{per_frame_eq_channel_dirch_form}. 

As such we can then employ a general complex basis expansion model (GCE-BEM), the defining difference of which is that we can leverage the inherent sensing implicit in OTFS communications to use $P$ basis functions with the true estimated Doppler shifts correspondent to our number of paths.
\[
  h(t)\;\approx\;\sum_{p=0}^{P-1}c_pe^{j2\pi\,f_p T_{s}}
\]
indexing the 2D OTFS channel from size $MNxMN$ we take the first column  due to the 2D circulant nature each column contains all unique from the overall matrix. We then treat each value as its own GCE-BEM estimator where the coefficients $c_p$ are reflective of the complex gain of the fractional component of each possible Doppler path. This then allows for us to utilize previous channel estimates to determine the phase rotation of each Doppler component across time.  A similar operation could be performed directly onto the DD pilot estimation grid of a single spike pilot which contains equal information content. 

The implicit strength in the delay-Doppler domain to estimate paths and their respective fractional Doppler shifts then renders this much more suitable for use than in OFDM operations. Further the increased coherence of the delay-Doppler channel paired with the consistent spreading of each fractional path allows for the  implicit assumption in CE-BEM that $C_p$ is approximately constant across our averaging window to be more accurate than that of OFDM systems. 

Improvements in the native ability to estimate the resultant OTFS channel during a high mobility scenario then further improves the ability for GCE-BEM estimation in the delay-Doppler domain when compared to the time-frequency. 

\subsubsection{Fractional OTFS Prediction Margin of Error}

GCE-BEM for best performance matches the basis function frequency directly to the true frequency of each channel path, as this is not possible with perfect accuracy we consider an extension which allows for iterative refinement of Doppler estimated $\nu$ to account for errors in frequency estimation from fractional channel estimation.

As GCE-BEM is a model based structure a number of iterative refinement algorithms can be applied, for this work we utilize basic Gauss Newton method optimization to account for small errors in fractional Doppler estimation. 
The GCE-BEM provides a coarse delay-Doppler map from which we extract an initial set of fractional Doppler estimates $\nu_0$ and corresponding complex path gains $\mathbf{C}_0$. To refine the Doppler values, we employ  \textbf{Gauss-Newton (GN)} iteration.

At each GN step $i$, the dictionary matrix $\mathbf{E}(\nu_i)\in\mathbb{C}^{T\times P}$ is built from the current Doppler vector $\nu_i$ and the centered symbol-time vector $\mathbf{t}\in\mathbb{R}^{T}$:
\begin{equation}
\mathbf{E}_{t,p}(\nu_i) = e^{\!\bigl(j 2\pi \nu_{i,p} \, t_t\bigr)},
\end{equation}
where $T$ is the number of time snapshots and $P$ is the number of paths.

The gains are first solved in closed form via regularized least squares:
\begin{equation}
\mathbf{C}_i = \bigl(\mathbf{E}^H\mathbf{E} + \lambda_{\text{amp}}\mathbf{I}_P\bigr)^{-1} \mathbf{E}^H\mathbf{H},
\end{equation}
with $\mathbf{H}\in\mathbb{C}^{T\times B}$ the observed history over $B$ beams and $\lambda_{\text{amp}}>0$ a small Tikhonov regularization.

The residual is computed as
\begin{equation}
\mathbf{R}_i = \mathbf{H} - \mathbf{E}(\nu_i)\mathbf{C}_i.
\end{equation}

The Jacobian with respect to Doppler is formed element-wise:
\begin{equation}
\mathbf{J}_i = j2\pi\,\mathbf{t}\,\mathbf{E}(\nu_i),
\end{equation}
where $\mathbf{t}$ is a column vector.

The approximate Hessian and gradient are
\begin{align}
\mathbf{H}_i &= \mathbf{J}_i^H\mathbf{J}_i, \\
\mathbf{g}_i &= \mathbf{J}_i^H\mathbf{R}_i.
\end{align}

With Levenberg--Marquardt damping $\lambda_{\text{GN}}$, we solve:
\begin{equation}
(\mathbf{H}_i + \lambda_{\text{GN}}\mathbf{I}_P)\,\Delta\nu_i = -\mathbf{g}_i
\end{equation}
and update the Doppler vector with a step size $\alpha \leq 1$:
\begin{equation}
\nu_{i+1} = \nu_i + \alpha\,\Delta\nu_i.
\end{equation}

The process typically converges in 5-10 iterations, yielding super-resolution Doppler estimates that are more robust to the fractional-bin errors inherent in the initial GCE-BEM output and to slow changes in Doppler frequency across OTFS frames. 

Although OTFS channel estimation is not the focus of this analysis it's important to note that the use superimposed or data aided virtual pilots may help facilitate accumulation the number of pilots required to estimate coefficients $c$ from frames within a channels given coherence window. \cite{OTFS_BEM_EST}. 

Such considerations must be made in tandem with the system design of the channel estimation and the mobility environment expected. 

\section{Simulation Environment \& Test Scenarios}
\label{se:Simulation}

\subsection{Channel Estimation \& Assumptions:}
\label{se:StaticChannel}

\subsubsection{Time-Frequency Channel Estimation}
In our OFDM benchmark, channel estimation is performed via a widely‐used \emph{scattered‐pilot} scheme across the time-frequency grid \cite{OFDM1}.  Pilot symbols \(P[n_p,k_p]\) are inserted at regular intervals of OFDM symbols in time and subcarriers in frequency, for multiple transmit antennas orthogonal symbols and positions are selected as to ensure clear estimation for each transmission antenna.  
\[
X[n,k] =
\begin{cases}
P[n_p,k_p], & (n,k)\in\mathcal P,\\
0, & \text{otherwise},
\end{cases}
\quad
\]
\[
\mathcal P = \bigl\{(n_p + i\Delta_n,\;k_p + j\Delta_k)\bigr\}\,.
\]
At each pilot location the least‐squares estimate of the channel is simply
\[
\hat H[n_p,k_p]
=\frac{Y[n_p,k_p]}{P[n_p,k_p]},
\]
where \(Y[n,k]\) is the received symbol on subcarrier \(k\) of OFDM symbol \(n\).  To obtain estimates at the data positions \((n,k)\notin\mathcal P\), we apply a two‐dimensional cubic interpolation over the pilot grid \cite{OFDM2}:
\[
\hat H[n,k]
=\mathrm{Interp}_{\rm cubic}\!\bigl\{\hat H[n_p,k_p]\bigr\},
\]
which fits piecewise third‐order polynomials in both the time and frequency dimensions.  Two-dimensional cubic interpolation provides a good estimation accuracy under moderate Doppler and frequency‐selective fading without a reliance on long term channel statistics such as covariance, making it a standard choice in OFDM simulations.  

\begin{figure}[h!]
\centering
    \includegraphics[width=1\columnwidth]{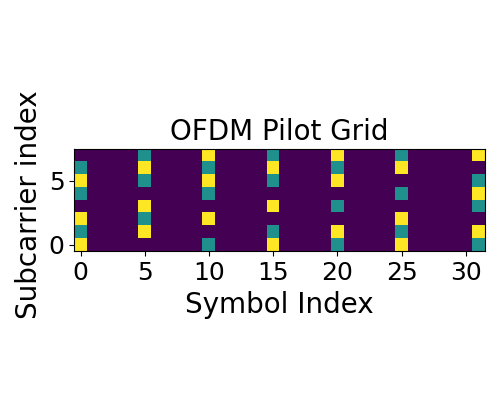}
    \caption{Time-Frequency Grid showing per antenna port scattered pilots}  
  \label{Fig.ScatteredPilotGrid}
\end{figure}

\subsubsection{Integer \& Fractional delay-Doppler Estimation}

In an idealized scenario where each path’s delay \(\tau_p\) and Doppler \(\nu_p\) align exactly to integer grid points \((\ell_p,k_p)\), the received pilot map \(Y_p[k,\ell]\) returns a clean impulse at \((k_p,\ell_p)\) and zeros elsewhere.  Accordingly, throughout our simulations under integer‐delay‐and‐Doppler assumptions we employ a simple least‐squares (LS) impulse‐based channel estimation in the delay-Doppler domain.

When fractional offsets arise, each path’s energy spreads into neighboring bins according to the Dirichlet kernel as shown previously. A variety of fractional‐tap estimation methods exist-e.g.\ weighted LS fitting of the Dirichlet shape over a local \((k,\ell)\) window, DFT oversampled impulse pilots, or pulse‐shaping designs with ambiguity‐function sidelobe suppression-but their exploration and evaluation is beyond the scope of this work.  Instead, to isolate and evaluate solely the performance of our precoder‐prediction framework under fractional Doppler, we assume \emph{ideal} channel estimation (i.e.\ perfect knowledge of \(\{\tau_p,\nu_p\}\)) whenever fractional delay-Doppler effects are considered. To maintain fair comparison we then consider ideal knowledge of subcarrier gains for OFDM systems in cases of fractional channels. 

\begin{figure}[h!]
\centering
    \includegraphics[width=1\columnwidth]{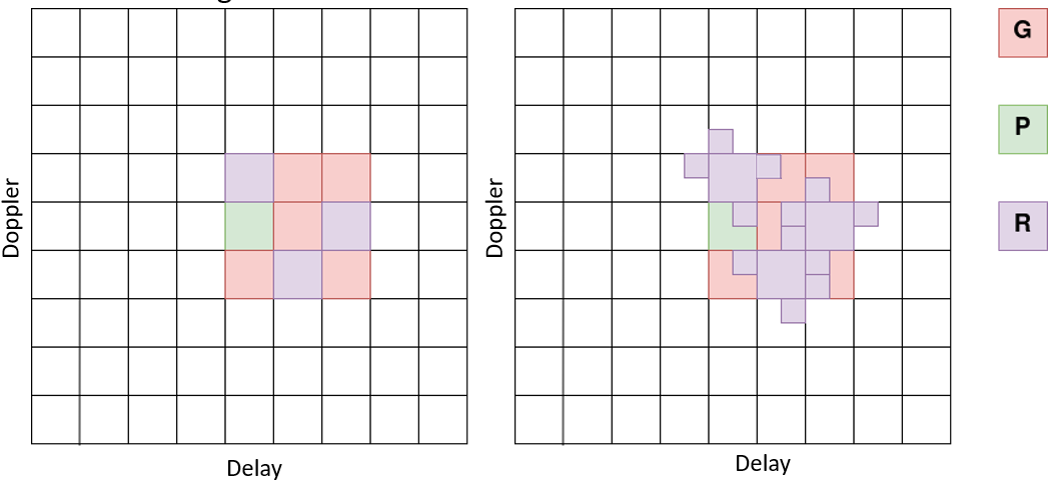}
    \caption{Delay-Doppler grid illustration of impulse‐based channel estimation Pilot (Green), Guard band (Red), Received channel taps (Purple)}  
  \textbf{Left:} Integer‐aligned paths.  
  \textbf{Right:} Fractional delay-Doppler offsets.
  \label{Fig.IntegerFractionalChannel_2}
\end{figure}

\subsubsection{Channel Assumptions}

\textit{Pilot Power and Parity:}
The total number and power of OFDM pilots used is assigned to be equal to that of the total number of pilots + guard band for OTFS estimation. The total power  pilots is balanced across modulations. 

\textit{CSI Feedback \& Precoding:}
Correspondent to the larger quantity of data sent by an OTFS frame there is thus an implicit limitation on feedback frequency, as such the maximum feedback frequency for our simulation is one frame of CSI delay. 
Correspondent with our unified data frame format, OTFS and OFDM systems will provide feedback at each data frame for use in the proceeding data frame. 

\textit{Environmental Ray Tracing:}
To evaluate the realism of our assumptions of the underlying channel conditions for each of our test scenarios we utilized MATLAB’s Ray Tracing Toolbox.  Through publicly available 3D geometry for buildings and terrain we test an urban and rural environment. We define a static transmitter and a mobile receiver. We then track the number of dominant paths as well as other corresponding signal components across a tenth of a second simulation window sampling positions every 1 ms. Mobile receivers followed pre-defined trajectories and diffraction, and specular reflections up to third-order were considered. 

We have then determined that practical channels consistent with our test cases frequently consist of  3 -5 prominent channel paths and that these paths are on average consistent across several milliseconds of time which is then greater than the duration of our data frames. As such our channels can safely be designed to accommodate 3-5 paths in a block fading model. To approximate change between frames we then test both static Doppler cases for integer channels and linear Doppler shifts for fractional channels.

\begin{figure}[h!]
\centering
    \includegraphics[width=0.8\columnwidth]{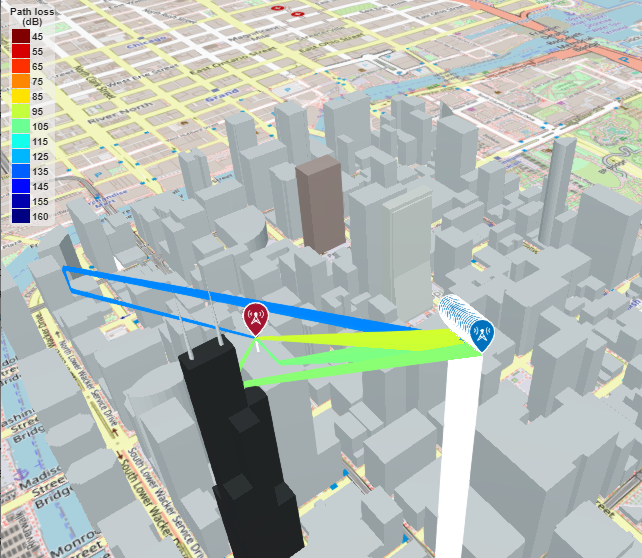}
    \caption{\textit{FANET} Ray Tracing(Urban)}
    \label{Fig.RayTrace}
\end{figure}

As shown in Fig.\ref{Fig.RayTrace} we can see that a mobile radio unit with mobility similar to that of a \textit{FANET} drone may operate in urban environments above the height of most buildings and as such receive 3 to 5 dominant paths. 

\textit{CSI Coherence}
As in implicit in Eq:\eqref{discrete_time_scattering_model} and supported by Fig. \ref{Fig.RayTrace}, we assume that each path’s complex gain, delay, and Doppler shift remain fixed over the duration of one OTFS frame. Although other works adopt a random-walk evolution for path attributes \cite{OTFS_ML_Pred}, these perturbations are minor in comparison to the dominant integer components over small sets of data frames and are not supported by the true environment. As such we assert that for all integer channels parameters are constant across OTFS frames.  Fractional channels are then tested for both static and linearly varying Doppler shifts representative of Doppler changing between frames corresponding with the angle of a reflected path.

 \subsection{Frame Structure:}
 For the purpose of maintaining parity in data transmission comparisons, an OTFS frame is defined to contain data of size  $\textit{M} \times \textit{N}$, where \textit{M} represents the number of Delay Bins and \textit{N} signifies the number of Doppler bins. Correspondingly, an OFDM frame is structured to comprise \textit{M} symbols, each utilizing \textit{N} subcarriers. This configuration ensures that both OFDM and OTFS frames encapsulate an identical quantity of data, thereby facilitating a direct and equitable comparison between the two modulation techniques. To match OTFS delay–Doppler bins, an equivalent OFDM frame uses $N$ subcarriers over $M$ symbols. This methodology is visualized below in Fig. \ref{Fig.Frame_Structure}

To further align the comparison and account for disparities in transmission characteristics, we adjust the cyclic prefix length for both OTFS and OFDM systems such that they are equal in overall length. 

\begin{figure}[h!]
\centering
    \includegraphics[width=0.8\columnwidth]{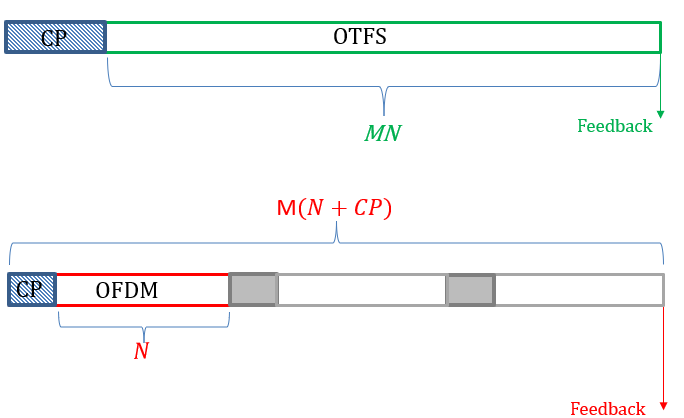}
    \caption{Transmit Frame Structure}
    \label{Fig.Frame_Structure}
\end{figure}

\textit{Test Scenarios:}
\begin{figure}[h!]
\centering
    \includegraphics[width=0.8\columnwidth]{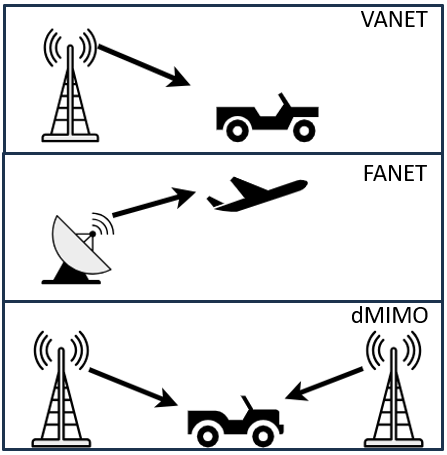}
    \caption{Simulation Test Scenarios}
    \label{Fig.TestScenarios}
\end{figure}
We utilize Monte Carlo simulation wherein a 50 channel realizations each with the same number of paths, and maximum delay-Doppler values is generated. Each test is then allowed to run for a duration that encompasses a startup time to acquire CSI history and a 10 transmit frame of transmission window during which metrics are taken. 

 Each scenario then varies by the selection of \textit{N} OTFS Doppler bins and \textit{M} OTFS delay bins, the use of integer or fractional channels and the selection of static or linearly shifting Doppler values. Each scenario involves a full rank MIMO transmission using QPSK modulation with an LMMSE receiver for equalization with hard decision decoding.

All simulations assume a carrier frequency of $3.5$ GHz, an OFDM subcarrier spacing of $15$kHz, and QPSK modulation with 3 paths.  
We evaluate three representative mobility and operational cases:

\begin{itemize}
  \item \textbf{\textit{VANET}}: a $2\times2$ MIMO Vehicular Ad‐Hoc Network, with mobile nodes traveling up to  $75 kph$, corresponding to a maximum Doppler shift of $225Hz$
  \item \textbf{\textit{FANET}}: a $2\times2$ MIMO Flying Ad‐Hoc Network (airborne), with speeds up to $144 kph$, yielding a maximum Doppler shift of approximately $450 Hz$.
  \item \textbf{\textit{dMIMO}}: a distributed MIMO scenario in which a mobile node (up to $75 kph$, $f_D\approx225\,$Hz) is served jointly by four transmit antennas representative of two transmit devices and two receive antennas (i.e.\ a $4\times2$ MIMO configuration). To account for the additional feedback overhead from two transmitters, we allow for 2 frames of feedback delay. 
\end{itemize}

\section{Results \& Discussion}
\label{se:Results}

\subsection{Integer Channel}

\begin{figure}[h!]
\centering
    \includegraphics[width=.8\columnwidth]{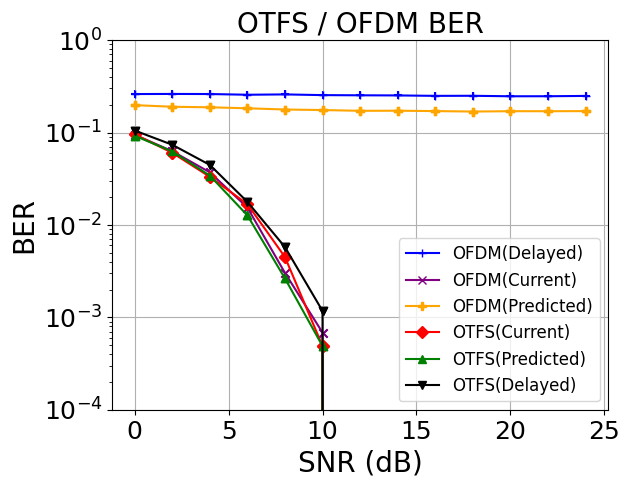}
    \caption{ Ideal Channel Estimation, \textit{VANET}, Integer(\(\pm225{}\)\,Hz)}
    \label{Fig.VANET_Ideal}
\end{figure}

\begin{figure}[h!]
\centering
    \includegraphics[width=.8\columnwidth]{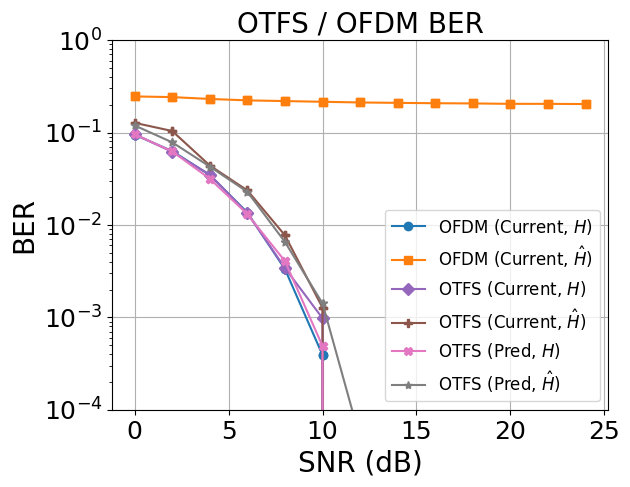}
    \caption{ Ideal/ Estimated Channel Comparison,Integer(\(\pm225{}\)\,Hz)}
    \label{Fig.H_EST_COMP}
\end{figure}

Integer channels utilize $8 \times 64$ delay-Doppler grid with 3 paths. Delay, Doppler and channel gains are considered static throughout each channel realization. Integer results can be differentiated into two types: Fig. \ref{Fig.VANET_Ideal}, shows the performance of integer prediction considering idealized channel estimation. We then show a performance comparison of how the OFDM and OTFS performance degrades when switching to estimated channel knowledge in \ref{Fig.H_EST_COMP}. For integer \textit{VANET} scenarios a basis function offset is employed to ensure the CE-BEM basis functions do not lie directly onto integer Doppler values. This is to prevent accidentally mapping  basis functions directly to all available Doppler values in the case of small integer restricted Doppler spread present in the \textit{VANET} scenario (e.g., \textit{VANET}: $\nu\in{-1,0,1}$; $f_q\in{-1,-\frac{2}{3},-\frac{1}{3},0,\frac{1}{3},\frac{2}{3},1}$)

The results show that, although ideal OTFS and OFDM perform similarly, ICI and mobility prevent OFDM prediction from accurately forecasting the channel. In contrast, OTFS maintains near-ideal performance under both ideal and estimated channel knowledge. It is important to note that although each scenario utilizes equal Doppler shifts that the inclusion of fractional values between the integer bounds reduces the effective Doppler experienced by OFDM, improving its fractional-case prediction.

\subsection{Fractional Channels}
\begin{figure}[h!]
\centering
    \includegraphics[width=.8\columnwidth]{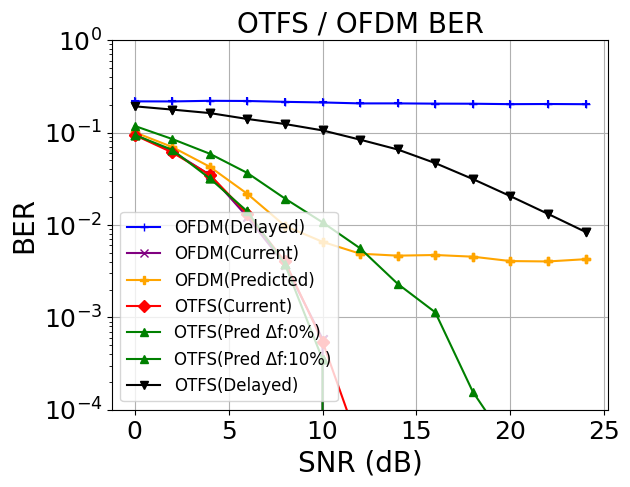}
    \caption{ Ideal Channel Estimation, \textit{VANET}, Fractional(\(\pm225{}\)\,Hz)}
    \label{Fig.FRAC_VANET}
\end{figure}

\begin{figure}[h!]
\centering
    \includegraphics[width=.8\columnwidth]{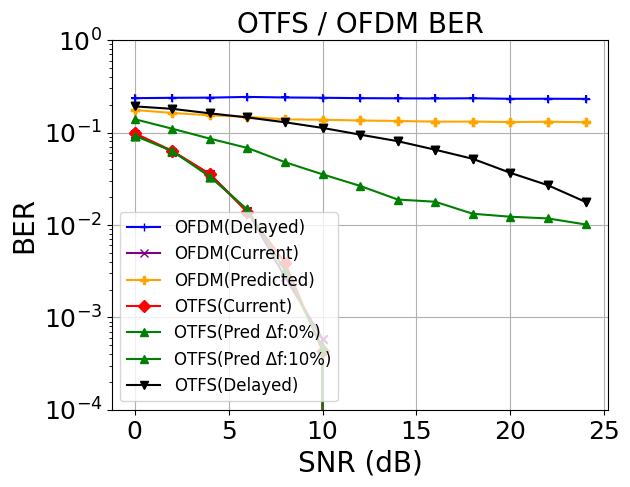}
    \caption{ Ideal Channel Estimation, \textit{FANET}, Fractional(\(\pm450{}\)\,Hz)}
    \label{Fig.FRAC_FANET}
\end{figure}

\begin{figure}[h!]
\centering
    \includegraphics[width=.8\columnwidth]{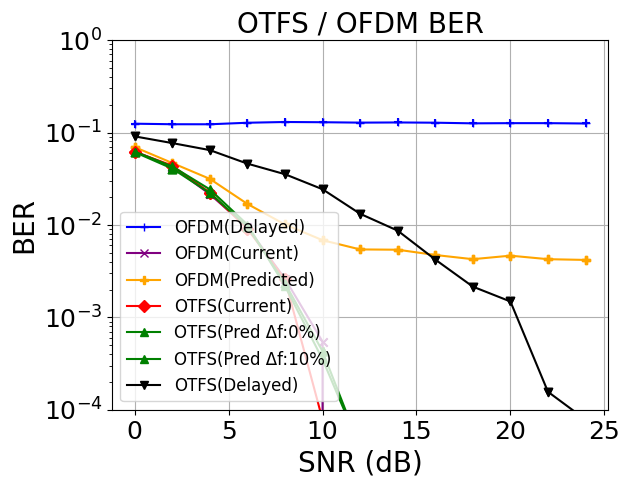}
    \caption{ Ideal Channel Estimation, \textit{dMIMO},Fractional(\(\pm225{}\)\,Hz)}
    \label{Fig.FRAC_dMIMO}
\end{figure}

\begin{figure}[h!]
\centering
    \includegraphics[width=.8\columnwidth]{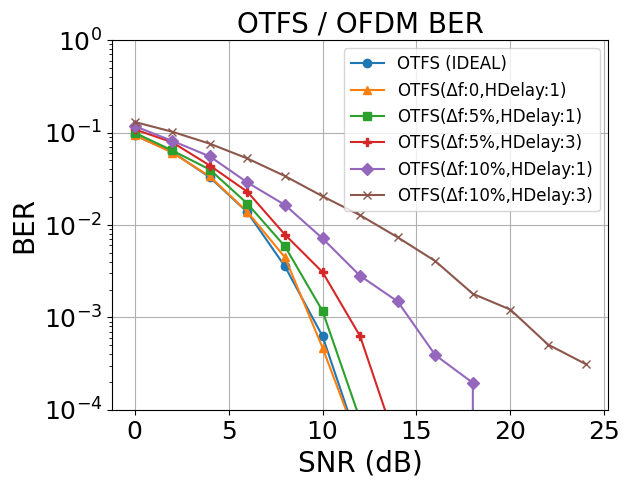}
    \caption{ Ideal Channel Estimation, Feedback Delay Evaluation, Fractional(\(\pm225{}\)\,Hz)}
    \label{Fig.FRAC_Feedback}
\end{figure}

Fractional scenarios consider an $8\times32$ delay-Doppler grid which maintains an equal effective Doppler frequency as the previous $8\times64$ size. Fractional results present of performance across \textit{VANET, FANET, dMIMO} scenarios wherein it is demonstrated that despite close fractional paths the native frequency estimation offered by OTFS path estimation provides a benefit for GCE-BEM based prediction. 
We additionally consider a static frequency estimation error $\Delta f$ of $10\%$ to account for difficulties in determining the exact frequencies of fractional Doppler paths Fig. \ref{Fig.FRAC_VANET},  Fig.\ref{Fig.FRAC_FANET},  Fig.\ref{Fig.FRAC_dMIMO} then demonstrates how the increased resilience afforded by OTFS prediction allows the increased overhead of distributed MIMO to be mitigated, improving overall system performance. Building upon this we demonstrate the ability of OTFS GCE-BEM to predict channels over greater feedback delays, demonstrating the performance of ideal channel knowledge, a variety of feedback delays and across multiple estimation error margins Fig. \ref{Fig.FRAC_Feedback}. 

\subsection{Considerations on Computational Complexity}
Traditional OTFS operations such as zero-forcing (ZF) precoding require inverting the full, vectorized OTFS channel matrix, which incurs a complexity of order \(\mathcal{O}(N M^3)\) per antenna pair. By predicting and updating the linear precoder forward in time, we eliminate the need to perform this expensive matrix inversion on each data frame, thereby achieving substantial computational savings.


\section{Conclusion and Open Challenges}
\label{se:conclusion}
In this study, we derived expressions for time-varying OTFS and OFDM I/O models, introducing an innovative method for predicting linear precoders through the interaction between OTFS modulation and the wireless channel. We then expand these to a basis expansion based prediction to offer performance improvement in fractional Doppler environments considering Doppler shift and inaccuracy in Doppler frequency estimation. 
We conduct a comprehensive analytical comparison to evaluate the impact of mobility on these systems, highlighting the enhanced predictability offered by OTFS modulation in dynamic channel environments. Our findings demonstrate that OTFS-based prediction techniques can significantly reduce feedback overhead through reduction in feedback frequency while improving system performance in cases where OFDM performance collapses. Through OTFS modulation and prediction, high-Doppler scenarios can be handled with reduced feedback and improved performance.

\bibliographystyle{IEEEtran}
\bibliography{References}{}

\end{document}